\documentclass[11pt]{article}

\newenvironment{proof}{\noindent {\em Proof}.\ }{\hspace*{\fill}$\halmos$\medskip}

\usepackage{graphicx,epsf,latexsym,amssymb}
\usepackage[mathscr]{eucal} 

\newcommand{\comment}[1]{}
\newcommand{\edo}{\end{document}}

\usepackage{amsmath}

\newcommand{\R}{{\mathbb R}}

\newcommand{\N}{{\mathbb N}}

\newtheorem{theorem}{Theorem}
\newtheorem{itlemma}{Lemma}[section] 

\newtheorem{itproposition}[itlemma]{Proposition}
\newtheorem{itcorollary}[itlemma]{Corollary}
\newtheorem{itremark}[itlemma]{Remark}
\newtheorem{itdefinition}[itlemma]{Definition}
\newtheorem{itexample}[itlemma]{Example}

\newenvironment{lemma}{\begin{itlemma}\rm}{\end{itlemma}} 
\newenvironment{remark}{\begin{itremark}\rm}{\end{itremark}} 
\newenvironment{corollary}{\begin{itcorollary}\rm}{\end{itcorollary}}
\newenvironment{proposition}{\begin{itproposition}\rm}{\end{itproposition}}
\newenvironment{definition}{\begin{itdefinition}\rm}{\end{itdefinition}}
\newenvironment{example}{\begin{itexample}\rm}{\end{itexample}}

\newcommand{\be}[1]{\begin{equation}\label{#1}}
\newcommand{\ee}{\end{equation}}
\newcommand{\bl}[1]{\begin{lemma}\label{#1}}
\newcommand{\ble}[1]{\begin{lemmaex}\label{#1}}
\newcommand{\br}[1]{\begin{remark}\label{#1}}
\newcommand{\bt}[1]{\begin{theorem}\label{#1}}
\newcommand{\bd}[1]{\begin{definition}\label{#1}}
\newcommand{\bp}[1]{\begin{proposition}\label{#1}}
\newcommand{\bc}[1]{\begin{corollary}\label{#1}}
\newcommand{\bfact}[1]{\begin{fact}\label{#1}}
\newcommand{\ber}[1]{\begin{exercise}\label{#1}}
\newcommand{\bex}[1]{\begin{example}\label{#1}}
\newcommand{\bem}[1]{\begin{example}\label{#1}}  
\newcommand{\ec}{\mybox\end{corollary}}
\newcommand{\efact}{\mybox\end{fact}}
\newcommand{\eer}{\mybox\end{exercise}}
\newcommand{\eex}{\mybox\end{example}}
\newcommand{\eem}{\mybox\end{example}}
\newcommand{\el}{\mybox\end{lemma}}
\newcommand{\ele}{\mybox\end{lemmaex}}
\newcommand{\er}{\mybox\end{remark}}
\newcommand{\et}{\qed\end{theorem}}
\newcommand{\ed}{\mybox\end{definition}}
\newcommand{\ep}{\mybox\end{proposition}}
\newcommand{\epr}{\end{proof}}
\newcommand{\bpr}{\begin{proof}}

\newcommand{\ecs}{\end{corollary}}
\newcommand{\eers}{\end{exercise}}
\newcommand{\eexs}{\end{example}}
\newcommand{\eems}{\end{example}}
\newcommand{\els}{\end{lemma}}
\newcommand{\eles}{\end{lemmaex}}
\newcommand{\ers}{\end{remark}}
\newcommand{\ets}{\end{theorem}}
\newcommand{\eds}{\end{definition}}
\newcommand{\eps}{\end{proposition}}
\newcommand{\halmos}{\rule{1ex}{1.4ex}}
\newcommand{\qed}{\hfill \halmos} 
\newcommand{\mybox}{\hfill $\Box$} 
\newcommand{\beq}{\begin{eqnarray}}
\newcommand{\eeq}{\end{eqnarray}}
\newcommand{\beqn}{\begin{eqnarray*}}
\newcommand{\eeqn}{\end{eqnarray*}}
\newcommand{\bi}{\begin{itemize}}
\newcommand{\ei}{\end{itemize}}
\newcommand{\ben}{\begin{enumerate}}
\newcommand{\een}{\end{enumerate}}

\newcommand{\col}{{\rm col}}

\newcommand{\bs}{\begin{split}}
\newcommand{\es}{\end{split}}

\newcommand{\cper}{conditionally persistent}
\newcommand{\bper}{bounded-persistent}

\newcommand{\sink}{sink}
\newcommand{\defeq}{\doteq}

\newcommand{\conservative}{stoichiometrically constrained}

\newcommand{\tilR}{\widetilde {R}}

\title{On persistence and cascade decompositions of\\
chemical reaction networks}
\author{David Angeli\thanks{Email: angeli@dsi.unifi.it, d.angeli@imperial.ac.uk,  Dept. of Electrical and Electronic Engineering, Imperial College, Dip. di Sistemi e Informatica, University of Firenze},
Patrick De Leenheer\thanks{
Email: deleenhe@math.ufl.edu. Supported in part by NSF Grant DMS-0614651. 
Dep. of Mathematics, University of Florida,
Gainesville, FL } and
 Eduardo D. Sontag\thanks{
Email:~sontag@math.rutgers.edu.
Supported in part by Grants NSF-0504557, NSF-0614371, and AFOSR FA9550-08. Dep. of Mathematics, Rutgers
University, Piscataway, NJ }
}
\begin{document}
\maketitle
\begin{abstract}
New checkable criteria for persistence of chemical reaction networks are
proposed, which extend and complement those obtained by the authors in
previous work.  The new results allow the consideration of reaction rates
which are time-varying, thus incorporating the effects of external signals,
and also relax the assumption of existence of global conservation laws, thus
allowing for inflows (production) and outflows (degradation). For
time-invariant networks parameter-dependent conditions for persistence of
certain classes of networks are provided.
As an illustration, two networks arising in the systems biology literature are
analyzed, namely a hypoxia and an apoptosis network. 
\end{abstract}
\section{Introduction}

For differential equations evolving in Euclidean space, ``persistence'' is the
property that all solutions starting in the positive orthant do not approach
the boundary of the orthant.  Interpreted for chemical reactions and
population models, this translates into a ``non-extinction property'' that
states that no species will tend to be completely eliminated in the course of
the reaction, provided that every species was present at the start of the
reaction. 

In the previous work~\cite{petri}, we presented criteria for checking
persistence in closed chemical reaction networks, couched in the language of
graph theory and Petri nets.  One of the main results was that a
time-invariant, conservative chemical network is persistent provided that each
siphon contains the support of a $P$-semiflow, regardless of the reaction
kinetics underlying the chemical reactions, or the values of parameters such
as rate constants.

In the present paper, we extend the previous results in several directions:
(1) Kinetic coefficients are now allowed to be time-varying.
(2) No conservation assumption is made (so solutions are potentially unbounded).
(3) The case in which there are critical siphons (i.e., do not contain the
support of any $P$-semiflow) is studied, and a sufficient condition for
persistence is provided in that case.
This latter condition is parameter-dependent.

The motivation for considering time-dependent coefficients is that these may
be used to represent the effect of external inputs to the network, while
not making a conservation hypothesis allows the consideration of inflows and
outflows, or production and degradation processes.
Finally, critical siphons arise in many examples.

We provide two examples of applications of the new results.  The first one
analyzes a model of the common core subsystem responsible for the hypoxia
control network in \emph{C.elegans}, \emph{Drosophila}, and humans.  Hypoxia
(deprivation of adequate oxygen supply) results in the expression of specific
genes in response to stress caused by low concentration of available oxygen.
This particular example was picked for two reasons.  First of all, viewing
oxygen concentration as an external input gives rise to a network with
time-dependent kinetic coefficients.  Second, in this model there are no
conservation laws that guarantee boundedness of solutions, and so the same
example serves to illustrate the role of the new concepts of conditional
persistence introduced in this work.  
The second example is an apoptosis (programmed cell death) network.
We characterize persistence using our result for critical siphons.

\section{Background on chemical reaction networks}
\label{Chemical Networks}

A \emph{chemical reaction network} (``CRN'', for short) is a list of chemical
reactions $R_i$, taking place among species $S_j$, where the indices
$i$ and $j$ take values in 
$\mathcal{R}:=\{1,2, \ldots , n_r \}$ and  
$\mathcal{S} := \{1,2, \ldots n_s \}$ respectively.
Individual reactions are then denoted as follows: 
\begin{equation}
\label{list}
 R_i: \quad \sum_{j \in \mathcal{S} } \alpha _{ij} S_j \rightarrow \sum_{j \in \mathcal{S} } \beta _{ij} S_j 
\end{equation}
where the $\alpha _{ij}$ and $\beta _{ij}$ are nonnegative integers called the 
\emph{stoichiometry coefficients}.
The species $j$ on the left-hand side for which $\alpha _{ij}>0$ are called 
\emph{reactants} and the ones on the right-hand side for which $\beta >0$ the
\emph{products}, of the reaction. 
Informally speaking, the forward arrow means that the transformation of
reactants into products only happens  in the direction of the arrow. If also
the converse transformation occurs, then, the reaction is 
reversible and we need to also list its inverse in the 
network as a separate reaction.
Sometimes, for convenience of notation, we will associate to a reaction the
two integer column vectors $\alpha _i$ and $\beta _i$, whose 
entries are defined by the stoichiometry coefficients. 
It is worth pointing out that we allow chemical reactions in which both the right and left hand sides are actually empty (though not at the same time).
This case corresponds, from a physical point of view, to \emph{inflows} and
\emph{outflows} of the chemical reaction.

As usually done, we arrange the stoichiometry coefficients into an $n_s\times 
n_r$ matrix, called the \emph{stoichiometry matrix} $\Gamma $, 
defined as follows: 
\begin{equation}
\label{stocmatrix}
[\Gamma ]_{ji} = \beta _{ij}-\alpha _{ij},
\end{equation}
for all $i \in \mathcal{R}$ and all $j \in \mathcal{S}$ 
(notice the reversal of indices).
This will be later used in order to synthetically write the
differential equation associated to a given chemical network.
Notice that we allow $\Gamma $ to have columns which differ only by their sign;
this happens when there are reversible reactions in the network. 

We discuss, next, how the speed of reactions is affected by the concentrations
of the different species. 
Each chemical reaction takes place continuously in time, at its own rate,
which is assumed to be only a function of the concentration of the species
taking part in it. 
In order to make this more precise, we 
define the vector $S = [S_1, S_2, \ldots S_{n_s} ]^{\prime}$ of species 
concentrations and, as a function of it, the vector of reaction rates
\[
R(S,t):= [R_1 (S,t), R_2 (S,t), \ldots R_{n_r} (S,t)]^\prime \, 
\]
where $t\in[0,\infty)$ denotes time.
Notice that we explicitly allow time-dependence, as we wish to
to consider the effect of external inputs to the system.
Such inputs may represent chemical species which are not explicitly considered
as part of the state variables but which, nevertheless, influence the reaction
rates.

Some mild uniformity requirements are needed for technical reasons as far as
time time-dependence is concerned. 
We assume, in particular, that for all $i \in \mathcal{R}$ for all $S$ and all $t\geq 0$
\be{ratesass} 
R_i(0,t) = 0 \textrm{ and } \underline{R}_i(S) \leq R_i(S,t) \leq \bar{R}_i (S),
\ee 
where the $\underline{R}_i (S), \bar{R}_i (S)$ are non-negative,
continuous functions of $S$, satisfying the following monotonicity constraint:
\be{monotonelowerbound}
S \gg_{R_i} \hat S \, \Rightarrow \bar{R}_i (S) > \bar{R}_i (\hat S) 
\ee
for all $i \in \mathcal{R}$ (and a similarly for $\underline{R}_i$),
where the notation $S \gg_{R_i} \hat S$ means that we have a strict inequality
$S_j>\hat S_j$ whenever species $j$ is a reactant in reaction $i$.
(We also write, more generally, $S \gg \hat S$ for any two vectors of species
concentrations, if $S_j>\hat S_j$ for all $j=1,\ldots ,n_s$.)
 
Furthermore, we assume standard regularity assumptions of $R_i(S,t)$ in order to ensure local existence and uniqueness of solutions.

A special form of reaction rates are \emph{mass-action kinetics}, which
correspond to the following expression:  
\[
R_i(S,t) = k_i (t) \prod_{j=1}^{n_s} S_j^{\alpha _{ij} } 
\quad\quad\mbox{for all}\, i=1,\ldots ,n_r \,
\]
(interpreting $S^0=1$ for all $S$),
that is, the speed of each reaction is proportional to the concentration of
its reagents.
Notice that we allow a time-varying kinetic rate $k_i(t)$, which may
account for the effect of external species not explicitly included in the
network under consideration. 
In the case of mass-action kinetics, a uniform lower and upper bound on
$R_i(S,t)$ exists if and only if there exist constants
$k^i_{\inf} >0$ and $k^i_{\sup} > 0$ such that
\[
k^i_{\inf} \leq k(t) \leq k^i_{\sup} \quad\forall\,t\geq0 \,.
\]

With the above notations, the chemical reaction network can be
described by the following system of differential equations: 
\begin{equation}
\label{chemreactionnetwork}
 \dot {S} (t) = \Gamma \, R(S(t),t).
\end{equation}
where $S=S(t)$ evolves in $\R^n_{\geq 0}$ and represents the vector of all
species concentrations at time $t$, and $\Gamma $ is the stoichiometry matrix. 
For systems with mass-action kinetics the following alternative expression
is valid: 
\begin{equation}
\label{shortnotation}
  \dot {S} (t) =  \sum_{ i \in \mathcal{R} } ( \beta _i - \alpha _i ) k_i (t) S^{\alpha _i} (t)
\end{equation}
where $\beta _i=\beta _{i*}$ is the column vector $\col(\beta _{i1},\ldots ,\beta _{in_s})$,
      $\alpha _i=\alpha _{i*}$ is the column vector $\col(\alpha _{i1},\ldots ,\alpha _{in_s})$,
and $S^\gamma =S_1^{\gamma _1}\ldots S_{n_s}^{\gamma _{n_s}}$ for any nonnegative vector
$\gamma =(\gamma _1,\ldots ,\gamma _{n_s})$.

It is straightforward to verify that the positive orthant is positively
invariant for system~(\ref{chemreactionnetwork}). Moreover,
for each $S_0 \in \R^n_{\geq 0}$ the affine subspace defined by:
\[ S_0 + \textrm{Im}[ \Gamma ] \]
is also invariant, regardless of the specific expression of reaction rates; its intersection with the positive orthant (which is therefore a forward invariant set)   
is called the \emph{stoichiometry class} of $S_0$.

\section{Petri nets and structural invariants}
\label{Petri Nets}

In stating our results, we will employ some terminology borrowed from
the graph theory, and specifically Petri nets.
Although arising from the study of discrete processes, Petri nets provide a
useful language and graphical representation for CRN's, and a number of
structural and analytical tools developed for them can be easily adapted
to the continuous context of chemical reactions.
In what follows, we associate to a CRN a bipartite directed graph (i.e., a
directed graph with two types of nodes) with weighted edges, called the
\emph{species-reaction Petri net}, or SR-net for short.
Mathematically, this is a quadruple
\[
( V_S, V_R, E, W)\,,
\]
where $V_S$ is a finite set of nodes, each one associated to a species, $V_R$
similarly is a finite set of nodes (disjoint from $V_S$) corresponding to 
reactions, and $E$ is a set of edges as described below. 
(We often write $S$ or $V_S$ interchangeably, or $R$ instead of $V_R$, by
identifying species or reactions with their respective indexes; the context
should make the meaning clear.)
The set of all nodes is also denoted by $V \defeq V_R \cup V_S$.

The edge set $E\subset V \times V$ is defined as follows.
Whenever a certain reaction $R_i$ belongs to the CRN:
\begin{equation}
\label{arbitrarydirection}
 \sum_{j \in \mathcal{S} } \alpha _{ij} S_j \quad \rightarrow 
 \quad \sum_{j \in \mathcal{S} } \beta _{ij} S_j \,,
\end{equation}
we draw an edge from $S_j \in V_S$ to $R_i \in V_R$ for all 
$S_j$'s such that $\alpha _{ij} >0$.  That is, $(S_j,R_i) \in E$ iff
$\alpha _{ij}>0$, 
and we say in this case that $R_i$ is an \emph{output reaction for} $S_j$.
Similarly, we draw an edge from $R_i \in V_R$ to every
$S_j \in V_S$ such that $\beta _{ij}>0$.
That is, $(R_i,S_j) \in E$ whenever $\beta _{ij}>0$,
and we say in this case that $R_i$ is an \emph{input reaction for} $S_j$.

Notice that edges only connect species to reactions and vice versa, but never
connect two species or two reactions.

More generally, given a nonempty subset $\Sigma \subseteq {\cal S}$ of species, 
we say that a reaction $R_i$ is an \emph{output (input) reaction for} $\Sigma$ if it is an output (input) reaction 
to some species of $\Sigma$.

The last element to fully define the Petri net is the function
$W:E\rightarrow \N$, which associates to each edge   
a positive integer according to the rule:
\[
W( S_j,R_i ) = \alpha _{ij} \quad\mbox{ and }\quad W(R_i, S_j) = \beta _{ij} \,.
\]
The stoichiometry matrix $\Gamma $, previously introduced,  is usually referred to as 
\emph{incidence matrix} in the Petri Net literature.

Several other definitions which are commonly used in that context will be of interest in the following. 
We say that a row or column vector $v$ is non-negative, and we denote it by
$v \succeq 0$ if it is so entry-wise.
We write $v \succ 0$ if $v \succeq 0$ and $v \neq 0$. 
A stronger notion is instead $v \gg 0$, which indicates $v_i > 0$ for all $i$.

\bd{P-semiflow}
A \emph{$P$-semiflow} is a row vector $c \succ 0$ such that $c \,
\Gamma = 0$. 
The \emph{support} of a $P$-semiflow is the set of indexes 
$\{ i \in V_S : c_i > 0\}$.
\eds

Using the fact that the entries of $\Gamma $ are integers, it is easy to show
that, given any $P$-semiflow $c$, there is always a $P$-semiflow with integer
components which has the same support as $c$.

\bd{def:conservative}
A nonempty subset $\Sigma \subseteq {\cal S}$ of species is \emph{{\conservative}}
if there is a $P$-semiflow whose support is included in $\Sigma $.
When $\Sigma ={\cal S}$, that is, if there is some $P$-semiflow $c \gg 0$,
we simply say that the CRN (or the corresponding Petri net)
is {\conservative}.
\eds

$P$-semiflows for the system (\ref{chemreactionnetwork}) correspond to
non-negative linear first integrals, that is, linear functions $S \mapsto cS$
such that $(d/dt)cS(t) \equiv 0$ along all solutions
of~(\ref{chemreactionnetwork}). 
In particular, a Petri net is {\conservative} if and only if there is a positive
linear conserved quantity for the system.
(Petri net theory views Petri nets as ``token-passing'' systems, and, in
that context, $P$-semiflows, also called \emph{place-invariants}, amount
to conservation relations for the ``place markings'' of the network, that show
how many tokens there are in each ``place,'' the nodes associated to species
in SR-nets.  We do not make use of this interpretation in this paper.)

\bd{T-semiflow}
A \emph{$T$-semiflow} is a column vector $v \succ 0$ such that $\Gamma \, v=0$.
\eds

Once again, one can assume without loss of generality that such a $v$ has
integer entries. 

\bd{def:consistentR}
A nonempty subset $\Lambda \subseteq {\cal R}$ of reactions is \emph{consistent} if there is a
$T$-semiflow whose support includes $\Lambda $.
When $\Lambda ={\cal R}$, we also say that the CRN, or its associated Petri net, is
consistent. 
\eds

The notion of $T$-semiflow corresponds to the existence of a collection of
positive reaction rates which do not produce any variation in the
concentrations of the species.  In other words, $v$ can be viewed as a
set of \emph{fluxes} that is in equilibrium~\cite{schuster}.
(In Petri net theory, the terminology is ``T-invariant,'' and the fluxes are 
flows of tokens.)

A vector $v=(0,0,\ldots ,0,1,0,\ldots ,0)$ with a ``1'' in the $i$th position and
0's elsewhere represents the $i$th reaction; thus we may label such a unit
vector as ``$R_i$''.  With this notational convention, the following fact
holds.  Suppose that $R_k$ and $R_\ell$ are reactions that are reverses of
each other, that is, $\alpha _{kj}=\beta _{\ell j}$ and $\beta _{kj}=\alpha _{\ell j}$
for every species $j\in {\cal S}$.  Then, the vector $R_k+R_\ell$ is a $T$-semiflow,
becaus the $k$th and $\ell$th columns of $\Gamma $ are opposites of each other.
In chemical network models of biological systems, it is common for several
of the reactions to be considered as reversible.  This gives rise to many such
``trivial'' $T$-semiflows.

\bd{siphon}
A nonempty set $\Sigma \subset V_S$ is called a \emph{siphon} 
if each input reaction  
for $\Sigma $ is also an output reaction for $\Sigma $.
A siphon is \emph{minimal} if it does not contain (strictly) any other
siphons.
\eds
For later use, we associate a particular set to a siphon $\Sigma $ as follows:
$$
L_\Sigma =\{x\in \R_{\geq 0}^{n_s} \,|\, x_i=0 \Longleftrightarrow i\in\Sigma \}.
$$
The set $L_\Sigma $ is therefore characterized as the set of concentration
vectors whose entries are zero if (and only if) the corresponding chemical
species are in the siphon $\Sigma $.  

\comment{
It is also useful to introduce a binary relation ``reacts to'', which we denote by
$\rightarrow tail$, and we define as follows: $S_i \rightarrow tail S_j$
whenever there exists a chemical reaction $R_k$, so that 
\[
\sum_{l  \in \mathcal{S}} \alpha _{kl} S_l \rightarrow \sum_{l \in \mathcal{S}}
   \beta _{kl} S_l
\]
with $\alpha _{ki}>0$, $\beta _{kj}>0$.
If the reaction number is important, we also write
\[
S_i \rightarrow tail^{ \hspace{-3mm} k} S_j
\]
(where $k \in \mathcal{R}$).
With this notation, the notion of siphon can be rephrased as follows:
$Z \subset \mathcal{S}$ is a siphon for a chemical reaction network if
for every $S \in Z$ and $k \in \mathcal{R}$ such that 
$\tilde{S}_k := \{ T \in \mathcal{S} : T \rightarrow tail^{ \hspace{-3mm} k} S \} \neq \emptyset$,  
it holds $\tilde{S}_k  \cap Z \neq \emptyset$. }

\section{Persistence and consistency}

Our main interest is the study of persistence: when do species remain nonzero,
if they start nonzero?  We will study two variants of this concept, and will
provide a necessary characterization for one and a sufficient characterization
for the other.

\bd{def:persistentset}
A nonempty subset $\Sigma \subseteq {\cal S}$ of species is \emph{\cper} (respectively, {\bper})
if there exists a bounded solution $S(\cdot )$ with $S(0)\gg0$
such that
\be{eq:persistenttraj}
\liminf_{t\rightarrow \infty } S_j(t) > 0
\quad\forall\, j \in  \Sigma 
\ee
(respectively, if this property holds for all bounded solutions).
When $\Sigma ={\cal S}$, we say simply that the corresponding CRN is {\cper} or
{\bper} respectively.
\eds

Note that in case $\Sigma ={\cal S}$
condition~(\ref{eq:persistenttraj}) amounts to the requirement that
the omega-limit set $\omega (S(0))$ should not intersect the boundary of the main
orthant.

The following result generalizes Theorem~1 in~\cite{petri} to systems with
time-varying rates, and provides a necessary condition for persistence.
It is proved in Section~\ref{sec:keyresults}.

\bt{cor:wpconssystem}
Every {\cper} CRN is consistent.
\ets

The following result gives a sufficient condition for persistence, and
generalizes Theorem~2 in~\cite{petri} to systems with time-varying rates:
It is proved in Section~\ref{sec:keyresults}.

\bt{theo:conservativesyphons}
If a CNR has the property that every siphon is {\conservative}, then 
it is {\bper}.
\ets

\section{Key technical results}
\label{sec:keyresults}

\bd{def:wpR}
A nonempty subset $\Lambda \subseteq {\cal R}$ of reactions is \emph{\cper} 
if there exists a bounded solution $S(\cdot )$ with $S(0)\gg0$
such that
\[
\liminf_{t\rightarrow \infty } R_i(S(t),t) > 0
\quad\forall\, i \in  \Lambda \,.
\]
\eds

The first key technical fact that we need is as follows;
it is proved in Section~\ref{sec:proof1}.

\bt{theo:persistence}
Every {\cper} subset $\Lambda $ of reactions is consistent.
\ets

\bd{def:extinctionset}
A nonempty subset $\Sigma \subseteq {\cal S}$ of species is an \emph{extinction set}
if there exists a bounded solution $S(\cdot )$ with $S(0)\gg0$ and a sequence
$t_n\rightarrow \infty $ such that
\be{eq:extinctionset}
\lim_{n\rightarrow \infty } S_j(t_n) = 0
\quad\Leftrightarrow\quad
j \in  \Sigma \,.
\ee
\eds

Equivalently, $\Sigma $ is an extinction set if and only if 
$L_\Sigma \bigcap \omega (S(0))\not= \emptyset$ for some bounded solution $S(\cdot )$.

The second key technical fact, proved
in~Section~\ref{sec:proof2}, is as follows.

\bt{theo:extinction}
Every extinction set of species is a siphon.
\ets

\bl{lem:conservative}
An extinction set cannot be {\conservative}.
\els

\bpr
Let $\Sigma $ be an extinction set, and pick a bounded solution $S(\cdot )$ as in the
definition of extinction.
Suppose that $\Sigma \subseteq {\cal S}$ is {\conservative}.  Let $c$ be a $P$-semiflow
whose support is included in $\Sigma $.
Since $c$ is a $P$-semiflow, $cS(t)=cS(0)>0$ for all $t\geq 0$ (the last inequality
because $S(0)\gg0$ and $c>0$).
Since the support of $c$ is a subset of $\Sigma $, it follows that
$cS(t) = \sum c_j S_j(t)$, with the sum only over the indices $j\in \Sigma $.
Thus~(\ref{eq:extinctionset}) cannot hold.
\epr

We defined what it means for a reaction $R_i$ to be an output reaction for a
species $S_j$, namely that $S_j$ should be a reactant of $R_i$.  More generally, we use
the following concept.

\bd{def:reactionsdepend}
Consider a nonempty subset $\Sigma \subseteq {\cal S}$ of species.
A reaction $R_i$ is said to be 
a \emph{{\sink}} for $\Sigma $ if the set of reactants of $R_i$ is a
subset of $\Sigma $.
The set of all {\sink}s for $\Sigma $ is denoted as $\Lambda (\Sigma )$.
\eds

\bl{lem:wp}
If $\Sigma $ is {\cper}, then $\Lambda (\Sigma )$ is {\cper}.
\els

\bpr
Suppose that we have a bounded solution such that
$\liminf_{t\rightarrow \infty } S_j(t) \geq  s >0$ for every $j \in  \Sigma $.
Since the solution $S(\cdot )$ is bounded, this means its closure is a compact
subset $K$ of the (closed) positive orthant.
Pick any {\sink} $R_i$ for $\Sigma $.
By Property~(\ref{monotonelowerbound}),
$\underline{R}_i (S) > 0$ for all $S\in K$.
Therefore $\liminf_{t\rightarrow \infty } R_i(S(t),t) > 0$ for this same trajectory, which
proves that $\Lambda (\Sigma )$ is {\cper}.
\epr

\bc{cor:wpcons}
If $\Sigma $ is {\cper}, then $\Lambda (\Sigma )$ is consistent.
\ecs

\bpr
This follows immediately from
Lemma~\ref{lem:wp} and Theorem~\ref{theo:persistence}.
\epr

This completes the proof of Theorem~\ref{cor:wpconssystem}, because
the hypothesis of the Theorem says that $\Sigma ={\cal S}$ is {\cper}.
By Corollary~\ref{cor:wpcons}, $\Lambda ({\cal S})={\cal R}$ is consistent, which means that the
CRN is consistent, as claimed.

To prove Theorem~\ref{theo:conservativesyphons}, we observe:

\bl{lem:notpersistextinct}
If a CRN is not {\bper}, then there is some extinction set.
\els

\bpr
Suppose that there is some bounded solution $S(\cdot )$ with $S(0)\gg0$, some
species $j_0\in {\cal S}$, and some sequence $t_n\rightarrow \infty $, such that 
$\lim_{n\rightarrow \infty } S_j(t_n) = 0$.
For this solution, and for this same sequence $\{t_n\}$, let $\Sigma $ be defined
as the set of species $j\in {\cal S}$ such that $\lim_{n\rightarrow \infty } S_j(t_n) = 0$.
Since $j_0\in \Sigma $, $\Sigma $ is nonempty, and it is an extinction set by definition.
\epr

Now Theorem~\ref{theo:conservativesyphons} follows from:

\bc{cor:notperistnoncons}
If a CRN is not {\bper}, then there is a non-{\conservative} siphon.
\ecs

\bpr
Assume that the given CNR is not {\bper}.
By Lemma~\ref{lem:notpersistextinct}, there is an extinction set $\Sigma $.
By Theorem~\ref{theo:extinction}, $\Sigma $ is a siphon.
By Lemma~\ref{lem:conservative}, $\Sigma $ is not {\conservative}.
\epr

\section{Proof of Theorem~\protect{\ref{theo:extinction}}}
\label{sec:proof2}

Let the nonempty subset $\Sigma \subseteq {\cal S}$ of species be an extinction set.
Pick a bounded solution $S(\cdot )$ with $S(0)\gg0$
such that $L_\Sigma \bigcap \omega (S(0))\not= \emptyset$ .
We need to prove that $\Sigma $ is a siphon.

Assume that $y \in L_\Sigma \cap \omega (S(0))$ but that $\Sigma $ is not a siphon.
Hence, there exists a species $S_j \in \Sigma $ so that for at least one of its input
reactions $R_k$ and all of $R_k$'s reactant species $S_l$, it holds $y_l >0$.
By Property~(\ref{ratesass}), 
we have that
\[
R_k(y,t) \geq \underline{R}_k ( y ) \defeq \bar{r} > 0
\]
for some positive value $\bar{r}$ and all $t \geq t_0$.   

Therefore, since all output reactions of $S_j$ have zero rate at $y$ (no
matter what the value of $t$ is), and at least some incoming reaction is
strictly positive, it follows by continuity of each of the
$\underline{R}_k(S)$'s that there is some $\varepsilon >0$, so that: 
\[
\dot {S}_j (t) = [ \Gamma R( z(t), t ) ]_j   \geq \bar{r}/2 
\]
whenever $z(t) 
\in \mathcal{B}_{\varepsilon }(y) \defeq \{ z \succeq 0: |z-y| \leq \varepsilon \}$ and
$t\geq t_0$. 
Now, using the uniform upper bound 
$\bar{R}(S)=(\bar{R}_1(S),\ldots ,\bar{R}_r(S))$
and its continuity, we know
that there exists $M>0$ so that $| \Gamma R(z,t)| \leq M$
for all $z$ as before. Hence, 
\be{speed}
 |S(t_b) - S(t_a )| \;= \;
\left | \int_{t_a}^{t_b} \Gamma R(S(t),t) \, dt \right | \;\leq \;(t_b - t_a ) M 
\ee 
whenever $S(t)\in {\cal B}_\varepsilon (y) $ for $t \in [ t_a, t_b ]$.

Assume without loss of generality (choosing a smaller $\varepsilon $ if necessary)
that $\varepsilon $ is such that $S(0)\not\in {\cal B}_{\varepsilon }(y)$.
Consider now any partial trajectory crossing the boundary of 
${\cal B}_{\varepsilon }(y)$ at time $t_{\varepsilon }$, and hitting the boundary of ${\cal B}_{\varepsilon /2}(y)$
at time $t_{\varepsilon /2}$, where $t_{\varepsilon /2}$ is picked as the first time after
$t_{\varepsilon }$ when this happens. 
Notice that such a partial trajectory exists, because 
$S(0)\not\in {\cal B}_{\varepsilon }(y)$ and by our assumption that $y\in \omega (S(0))$.

Since 
$\dot {S}_j (t) \geq \bar{r}/2$ 
for all $t \geq t_0$ whenever 
$z(t)$
belongs to
$\mathcal{B}_{\varepsilon } (y)$, it follows that necessarily we must exit
$\mathcal{B}_{\varepsilon } (y)$ an infinite number of times, hence infinitely many such
partial trajectories exist.

By the estimate in (\ref{speed}), the time it takes to get from the boundary
of $\mathcal{B}_{\varepsilon } (y)$ to $\mathcal{B}_{\varepsilon /2} (y)$ is at least $\varepsilon /2 M$.
Moreover, since $\dot {S}_j (t) \geq \bar{r}/2$ we have:
\[ 
\begin{array}{rcl}
S_j (t_{\varepsilon /2}) \;=\; 
S_j (t_{\varepsilon } ) + \int_{t_{\varepsilon } }^{t_{\varepsilon /2}} \dot {S}_j (t) \, dt
& \geq& \\ \;S_j (t_{\varepsilon } ) + \varepsilon \bar{r} / 4 M  & \geq& \varepsilon \bar{r} / 4M. 
\end{array}
\]
Obviously, for $t \geq t_{\varepsilon /2}$, and as long as $S(t)\in \mathcal{B}_{\varepsilon }(y)$, we
also have: $S_j (t ) \geq S_j (t_{\varepsilon /2} ) \geq  \varepsilon \bar{r}/ 4M$. 
This shows indeed $y \notin \omega (S(0))$, contradicting our hypothesis. Hence,
$\Sigma $ must be a siphon.
\qed

\section{Proof of Theorem~\protect{\ref{theo:persistence}}}
\label{sec:proof1}

Suppose that $\Lambda \subseteq {\cal R}$ is a {\cper} set of reactions, and pick a
bounded solution $S(\cdot )$ with $S(0)\gg0$
such that $\liminf_{t\rightarrow \infty } R_i(S(t),t) > 0$ for each $i\in  \Lambda $.
We need to show that $\Lambda $ is consistent.

Pick an arbitrary $t \geq t_0$.  Clearly:
\be{totakelimit}
S ( t ,t_0, S_0  ) - S_0 \;=\; \int_{t_0}^t  \dot {S} (t) \, dt  
\;=\;  \Gamma  \int_{t_0}^t R (S(t),t) \, dt\,. 
\ee
Since $S(t,t_0, S_0)$ is bounded, so are $R(S(t),t)$ 
and its average
\[
\frac{1}{t} \int_{t_0}^{t} R (S(t),t) \, dt  \,.
\]
Hence, there exists a sequence $t_n \rightarrow + \infty$ such that 
\[
\frac{1}{t_n} \int_{t_0}^{t_n} R (S(t),t) \, dt
\]
also admits a limit $\tilR \succeq 0$ as $n \rightarrow + \infty$. 
Now taking limits along this subsequence in both sides of (\ref{totakelimit}),
after dividing by $t$ yields:
\be{almostthere}
0 \;=\;\lim_{n \rightarrow + \infty} \frac{1}{t_n} 
\left(S ( t_n ,t_0, S_0  ) - S_0\right)\; =\; \Gamma \, \tilR\,.
\ee
Moreover, for all $i \in \Lambda $ we have
\[
\liminf_{t \rightarrow + \infty} R_i (S(t),t) \;=\; r_i  \;>\; 0
\]
and hence there exists $T >0$ so that for all $t \geq T$, 
$R_i(S(t),t) \geq r_i /2$. 
Letting $r$ be the minimum of the $r_i$'s,
we conclude that:
\[ \begin{array}{rl}
\tilR_i \;= & \lim_{n \rightarrow + \infty} 
\frac{1}{t_n} \int_{t_0}^{t_n} R_j (S(t),t) \, dt  \\ & \geq \;
\lim\frac{1}{t_n}(t_n-t_0)\frac{r}{2}\;=\;
\frac{r}{2}  \;>\; 0\,.
\end{array}
\]
So, $\tilR$ is a $T$-semiflow $v$ whose support contains $\Lambda $. 
\qed

\section{An example: hypoxia network}
As discussed in the Introduction, we analyze a model of the hypoxia control
network.  Starting from the model given in~\cite{kohn} for the core subsystem
of the \emph{hypoxia} control network in \emph{C.elegans}, \emph{Drosophila},
and humans, with 23 species and 32 reactions, the authors of~\cite{yu} picked
a subsystem consisting of 13 species and 19 reactions which constitute the key
components explaining experimentally observed behaviors.  We analyze this
simplified model.

One of the species, $S_1$, which represents the transcription factor HIF$\alpha $,
is subject to production and degradation (or, in formal terms, ``inflows'' and
``outflows'').

The reactions are as follows:
\be{hypoxia}
\begin{array}{ccccc}
0 & \rightarrow & S_1 & \rightarrow & 0 \\
S_1 + S_7 & \leftrightarrow & S_2 \\
S_2+ S_4 & \leftrightarrow & S_3  \\
S_1 + S_6 & \leftrightarrow & S_8 & \rightarrow & S_6 + S_{11} \\
S_4 + S_{10} & \leftrightarrow & S_9 \\
S_{10} & \leftrightarrow & S_{11} + S_{7} \\
S_2 + S_6 & \leftrightarrow & S_5 & \rightarrow & S_6 + S_{10} \\
S_{11}+ S_{12} & \leftrightarrow & S_{13} & \rightarrow & S_{12},
\end{array} 
\ee
where the meaning of the various biochemical species is in Table $\ref{tabel}$.

\begin{table}
\centering
\begin{tabular}{|c|c|}
\hline
$S1$  &HIF$\alpha$\\
\hline
$S2$ &  HIF$\alpha$:ARNT\\
\hline
$S3$  & HIF$\alpha$:ARNT:HRE\\
\hline
$S4$  & HRE\\
\hline
$S5$  & HIF$\alpha$:ARNT:PHD\\
\hline
$S6$  & PHD\\
\hline
$S7$  & ARNT\\
\hline
$S8$  & HIF$\alpha$:PHD\\
\hline
$S9$  & HIF$\alpha$ OH:ARNT:HRE\\
\hline
$S10$  & HIF$\alpha$ OH:ARNT\\
\hline
$S11$  & HIF$\alpha$ OH\\
\hline
$S12$  & VHL\\
\hline
$S13$  & HIF$\alpha$ OH:VHL\\
\hline
\end{tabular}
\caption{The various species in the hypoxia network (\ref{hypoxia}).}
\label{tabel}
\end{table}

External oxygen affects the dynamics of the system by scaling the rate
constants for the reactions $S_8  \rightarrow  S_6 + S_{11}$ and
$S_5  \rightarrow  S_6 + S_{10}$.
Mathematically, this means that $k_i(t)$ is proportional to the oxygen
concentration (and hence is potentially time-dependent) for each of these two
reactions.

According to \cite{kohn}, 
when the oxygen level falls below a critical value, a sharp rise in HIF$\alpha$ is observed, 
while this protein is undetectable if the oxygen level is above the critical value. 
The modeling effort in \cite{kohn} and the analysis of the model in \cite{yu} were aimed at understanding this  
switch-like behavior. For simplicity, the oxygen level was kept constant, and the model was investigated 
over a range of values of this constant. Here we will investigate the persistence properties of this network under the assumption 
that oxygen levels are time-varying.

\begin{figure}
\begin{center}
\includegraphics[width=7.5cm]{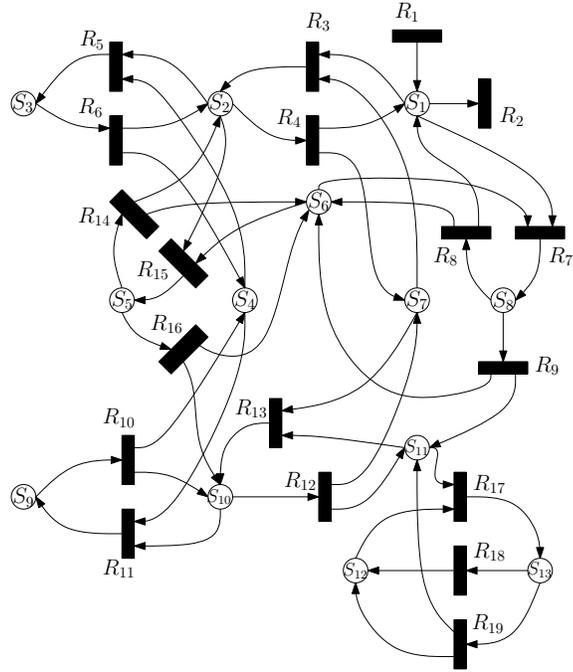}
\end{center}
\caption{Hypoxia network}
\label{noinputplace}
\end{figure}

The associated reaction network, represented as a Petri Net, is
shown in Fig.~\ref{noinputplace}.
This network admits $4$ linearly independent $P$-semiflows, which are
associated to the following conservation laws:
\be{Psemiflow}
\begin{array}{rcc}
S_{12} + S_{13} &=& \mbox{const}_1 \\
S_3 + S_4 + S_9 &=& \mbox{const}_2 \\
S_5 + S_6 + S_8 &=& \mbox{const}_3 \\
S_2 + S_3 + S_5 + S_7 + S_9 + S_{10} &=& \mbox{const}_4 
\end{array}
\ee
and it is clearly not conservative, due to the presence of outflows and
inflows (equation for $S_1$).
Not only are there no strictly positive conservation laws $v\Gamma =0$,
but there are not even ``decreasing'' semiflows satisfying $v\Gamma \leq 0$
which could be used as Lyapunov functions in order to establish boundedness
of solutions.
In addition, some kinetic rates are allowed to be time-varying.  
Thus, the techniques from~\cite{petri} cannot be applied to study persistence.

We consider next the possible $T$-semiflows.
There are several ``trivial'' ones, corresponding to the reversible
reactions:
$R_3 + R_4$, $R_5+R_6$, $R_7 + R_8$, $R_{10}+ R_{11}$, $R_{12}+
R_{13}$,$R_{14} + R_{15}$ and $R_{17} + R_{19}$.

In addition to these, one can find $3$ non-trivial independent $T$-semiflows:
\be{Tsemiflow} 
\begin{array}{l}
R_1 + R_2 \\
R_1 + R_7 + R_9 + R_{17} + R_{18} \\
R_{1} + R_3 + R_{15} + R_{16} + R_{12} + R_{17} + R_{18} \,.
\end{array}
\ee
Since every reaction appears in at least some $T$-semiflow, the sum of the
semiflows shown is also a semiflow which is strictly positive, and we can
conclude that the network is consistent.

Thus, the necessary condition for persistence in
Theorem~\ref{theo:persistence} is satisfied.  This does not quite prove
conditional persistence, but shows that the property is not ruled out by the
structure of the network.

Next, we find a set of minimal siphons:
\be{siphons}
\begin{array}{l}
\{ S_{12} , S_{13} \} \\
\{ S_3, S_4, S_9 \} \\
\{ S_5 , S_6, S_8 \} \\
\{ S_2,S_3,S_5,S_7,S_9,S_{10} \}.
\end{array}
\ee
Notice that all of them coincide with the support of some $P$-semiflow,
hence every siphon is {\conservative}.
We conclude that the network is {\bper}, by
Theorem~\ref{theo:conservativesyphons}.
The
next step is the investigation of which variables have the potential for becoming unbounded. 
An algorithm developed for this purpose is illustrated in \cite{cdcangeli}.
It is based on a linear time-varying embedding of individual species equations and it
carries out a consistency check in order to verify which scenarios are compatible with the
topology of the network, assuming (for instance) mass-action kinetics.
Scenarios are described by labeling each species with a symbol in $\{0,1,\omega \}$ depending on its asymptotic behaviour,
namely converging to $0$, bounded and bounded away of $0$ or diverging to infinity.
While such a classification of behaviours does not cover all the potential asymptotic dynamics of
general systems, it appears to be, in practical situations, a fairly mild restriction. 

\begin{figure}
\centerline{
\includegraphics[width=7cm]{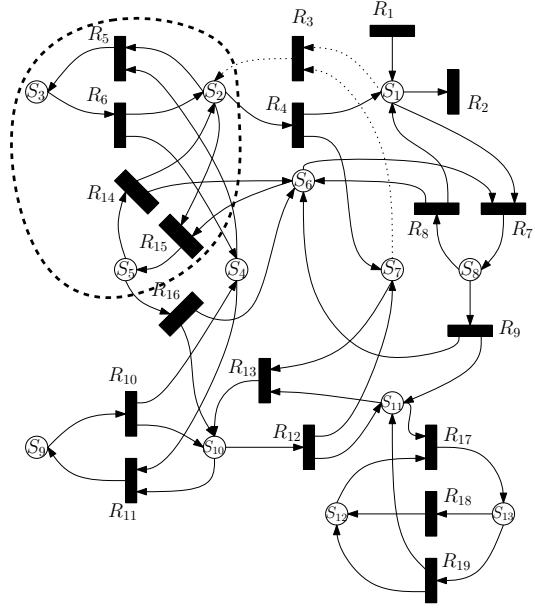} }
\caption{Reduced Petri Net and P-decreasing vector support}
\label{noinputflow}
\end{figure}

Running the algorithm on the hypoxia network 
yields $3$ potential scenarios for the asymptotic behaviour of the network: 
\begin{center}
\begin{tabular}{|c|ccccccc|}
\hline
Scenario & $S_1$ & $S_2$ & $S_3$ & $S_4$ & $S_5$ & $S_6$ & $S_7$  \\
\hline
I & 1 & 1 & 1 & 1 & 1 & 1 & 1  \\
II & 1 & 0 & 0 & 1 & 0 & 1 & 0  \\
III & 1 & 1 & 1 & 1 & 1 & 1 & 0   \\
\hline
 \hline
 & $S_8$ & $S_9$ & $S_{10}$ & $S_{11}$ & $S_{12}$ & $S_{13}$ & \\
 
 I & 1  & 1 & 1 &1 & 1 & 1 &  \\
 II & 1 & 1 & 1 & $\infty$ & 0 & 1 & \\
 III & 1 & 1 & 1 & $\infty$ & 0 & 1 & \\
 \hline
 \hline
\end{tabular}
\end{center}
\normalsize
Moreover, see \cite{cdcangeli} further details, asymptotic invariant
vector analysis allows to discard
Scenario III, see Fig.~\ref{noinputflow}). 
Indeed, the linear function $S_2 + S_3 + S_5$ is associated to a $P$-decreasing vector of the reduced net obtained by removing reaction $R_3$, which, according to the labeling is asymptotically switched off. 
Hence, in this scenario, the quantity $S_2 + S_3 + S_5$ gets asymptotically dissipated with a strictly positive rate; a clear contradiction.

Indeed, simulations showed that both scenarios I and II are possible for different values of the kinetic constants. Notice
that persistence is violated in scenario II as several species (namely $S_2, S_3, S_5, S_7, S_{12}$ ) vanish asymptotically.
This is not in contrast with our theoretical developments as species $S_{11}$ in this case gets accumulated and diverges
to infinity, thus violating the boundedness assumption which is crucial to Theorem \ref{theo:conservativesyphons}. This also shows that extinction sets for unbounded solutions need not be siphons.

\section{Cascade decompositions and siphons}

One of the main results in \cite{petri} states that a time-invariant, conservative CRN is persistent 
if each siphon contains the support of a $P$-semiflow, regardless of the 
reaction kinetics underlying the chemical reactions, or the values of 
parameters such as rate constants. In the previous sections we have extended 
this result to CRN's with time-varying reaction kinetics which may have inflows 
and outflows. In this section, we return to the time-independent case, and we explore 
the scenario of a CRN that has a siphon which does not contain the support of 
a $P$-semiflow (using the terminology of \cite{petri}, we say that this siphon is critical). 
We will see that a simple sufficient condition based on a linearization argument  
can be formulated that still guarantees persistence of the CRN.  
The linearization condition is related to, though different from, other
conditions imposed on chemical reaction networks in order to obtain
persistence, notably that given in Section 7.2.1 in~\cite{thesis:chaves}.

Let $\Sigma $ be a critical siphon. 
We partition the state $S$ according to species which do not belong to $\Sigma $, denoted by $\zeta $, and species which belong to $\Sigma $, denoted by
$\sigma $.
As shown in \cite{petri}, siphons have the forward invariance property stated below:
\[ S_0 \in L_{\Sigma} \Rightarrow S(t,S_0) \in L_{\Sigma} \; \forall \, t \geq 0 \]
If the system equations of the CRN are
\be{reactioncascade}
\left [ \begin{array}{c} \dot { \zeta } \\ \dot { \sigma } \end{array} \right ] \, = \, \left [
\begin{array}{c} f_{\zeta } ( \zeta , \sigma ) \\ f_{\sigma } ( \zeta , \sigma ) \end{array} \right ],
\ee 
then the forward invariance property implies the following condition:
\be{infinitesimal}
f_{\sigma } ( \zeta , 0) =0, \qquad \forall \zeta \succeq 0;
\ee
Consequently, the dynamics on the closure of $L_{\Sigma }$ (which is a forward invariant orthant of lower dimension, as proved in \cite{petri}) 
is completely determined by the equations:
\be{reducedzeta}
 \dot { \zeta } (t) = f_{\zeta } ( \zeta (t) , 0 ).
\ee
If $\omega (S_0) \cap L_{\Sigma }$ is non-empty for some $S_0 \gg 0$, then it is also a forward invariant set, and so is its closure, and the dynamics on the latter set are described by the 
evolution of the $\zeta $ components, as given by (\ref{reducedzeta}). 
Checking non-emptiness of such a set may in general be a rather challenging task. However, in some cases it turns out that this is indeed possible by analysis of a reduced variational equation. 
Since $\zeta $ in (\ref{infinitesimal}) is arbitrary, we have the following further result:
\be{differentialform}
\left . \frac{ \partial f_{\sigma } ( \zeta , \sigma ) }{ \partial \zeta } \right |_{\sigma =0} = 0, \qquad \forall 
\zeta \succeq 0.
\ee
Therefore, the variational equation associated to solutions corresponding to any initial condition $S(0)=[ \zeta (0)^{\prime}, 0 ]^{\prime}$, takes the following block triangular structure:
\be{triangular}
\left [\begin{array}{c} \dot { \delta \zeta } (t) \\ \dot { \delta \sigma } (t) \end{array} \right ] \, = \, \left [
\begin{array}{cc} \frac{ \partial f_{\zeta }  }{ \partial \zeta } ( \zeta (t),0 ) & \frac{ \partial f_{\zeta }  }{ \partial \sigma } (\zeta (t),0 )  \\ 0 & \frac{ \partial f_{\sigma }  }{ \partial \sigma } (\zeta (t),0) \end{array} \right ] \,
\left [\begin{array}{c} \delta \zeta  (t) \\  \delta \sigma  (t) \end{array} \right ].
\ee
These observations suggest that persistence may be understood by examining the stability properties of
a reduced order variational equation:
\be{reduced}
\dot { \delta \sigma } (t) = \frac{ \partial f_{\sigma } }{\partial \sigma } ( \zeta (t),0 ) \, \delta \sigma (t), 
\ee
whose dynamics may determine whether the part of the boundary where $\sigma=0$ is repelling or 
attracting to interior solutions. 
We claim that (\ref{reduced}) is a time-varying positive system because the matrix 
\be{metzler}
 \frac{ \partial f_{\sigma } }{\partial \sigma } ( \zeta ,0 )
\ee
is Metzler for all $\zeta \succeq 0$. This will follow from the forward invariance of the non-negative orthant for system (\ref{reactioncascade}).
Indeed, a standard first-order Taylor expansion yields:
\be{firstord}
\begin{array}{rcl}
f_{\sigma } ( \zeta , \sigma ) &=& f_{\sigma } ( \zeta ,0) + \frac{\partial f_{\sigma }}{ \partial \sigma } ( \zeta ,0) \sigma + o (| \sigma | ) \\
&=& \frac{\partial f_{\sigma }}{ \partial \sigma } ( \zeta ,0) \sigma + o (| \sigma | ).
\end{array}
\ee
If $[ \frac{\partial f_{\sigma }}{ \partial \sigma } ( \zeta ,0)]_{i,j}$ would be negative for some integers $i \neq j$ and some $\zeta $ and letting $\sigma = \varepsilon e_j$, ($e_j$ is the $j$-th element of the canonical basis of the Euclidean space of compatible dimension), yields, thanks to (\ref{firstord}), 
$f_{\sigma } ( \zeta , \varepsilon e_j )_i < 0$ for all sufficiently small $\varepsilon >0$. This violates forward invariance of the positive
orthant (since, clearly, $[\varepsilon e_j]_i=0$).

Assume that closed invariant sets of (\ref{reducedzeta}) are equilibria, let us denote them 
by $\zeta _e$. In view of what has been said so far, one is lead naturally to consider 
the following implications concerning persistence :
\begin{enumerate}
\item $\frac{ \partial f_{\sigma } }{\partial \sigma } ( \zeta _e,0 )$ is Hurwitz $\, \Rightarrow \, $ CRN is not persistent.
\item $\frac{ \partial f_{\sigma } }{\partial \sigma } ( \zeta _e,0 )$ has a positive dominant eigenvalue and is irreducible $ \, \Rightarrow \,$  
$\zeta _e \notin \omega (S_0)$,
\end{enumerate}
This suggests that persistence can sometimes 
be analyzed by simple linearization techniques around boundary equilibria. 
Item $1.$ of the above claim is obvious, as the existence of a boundary equilibrium point with a non-trivial stable manifold clearly violates persistence of a CRN.

We proceed next to a formal statement and proof of item $2$. The proof relies on a few auxiliary results which are deferred to the Appendix.
\bt{conservativenetworkpersistence}
Consider a time-invariant CRN whose associated Petri-Net is conservative. Let $\Sigma $ be a minimal siphon and let the state
$S(t)$ be partitioned accordingly: $[ \zeta (t), \sigma (t)]$. Let $\zeta _e$ be an equilibrium of
(\ref{reducedzeta}), globally asymptotically stable 
relative to its stoichiometry class. 
 Assume, moreover, that
\be{hyperbolic}
 \frac{\partial f_{\sigma } }{ \partial \sigma } ( \zeta _e , 0 ) \textrm{ be irreducible and } \lambda _{PF} \left ( \frac{\partial f_{\sigma } }{ \partial \sigma } ( \zeta _e , 0 ) \right ) > 0
 \ee
 Then, for any initial condition $S_0=[\zeta _0, \sigma _0]$ in the interior of the positive orthant and 
 denoting by $\omega (S_0)$ the $\omega $-limit set of the corresponding solution $S(t,S_0)$,
there holds that $S_e \doteq [\zeta _e,0] \notin \omega ( S_0)$.
\et
\bpr 
The proof is by contradiction. Let $S_0$ in the interior of the positive orthant be such that 
$S(t_n,S_0)\rightarrow S_e$ along some increasing sequence $t_n\rightarrow +\infty$ as $n\rightarrow +\infty$. 
Let $c \gg 0$ be a left eigenvector of $\frac{\partial f_{\sigma}}{\partial \sigma}(\zeta_e,0)$:
$$
c\frac{\partial f_{\sigma}}{\partial \sigma}(\zeta_e,0)=\lambda_{PF}\left(\frac{\partial f_{\sigma}}{\partial \sigma}(\zeta_e,0)\right)c
$$ 
and define the following function:
$$
V(S):=[0, c]S.
$$
Then $V(S_e)=0$ (since $S_e$ is of the form $[*,0]'$), and ${\dot V}(S)>0$ for all $S\in Q\cap \textrm{int}(\R^n_{\geq 0})$, 
where $Q$ is a neighborhood of $S_e$ chosen as in the proof of Lemma $\ref{aux1}$, applied 
to the vector function $f_\sigma(\zeta,\sigma)$. 
Furthermore, we assume without loss of generality that $Q$ is relatively open in $\R^n_{\geq 0}$. 
Returning to the solution starting at $S_0$, we claim the following:
\begin{center}
{\it There is some $p\in \omega(S_0)\cap L_{\Sigma}$ with $p\neq S_e$.}
\end{center}
By passing to a subsequence if necessary we assume that $S(t_n,S_0)\in Q$ for all $n$. 
There is some sufficiently large $n^*$ such that for all 
$n>n^*$, we can define
$$
\tau_n:=\sup \{t\, |\, t<t_n\textrm{ and } S(t,S_0)\notin Q\},
$$
which can be thought of as the most recent entry time (into $Q$) before time $t_n$. 
Notice that $n^*$ exists, so that the definition of the $\tau_n$'s makes sense. 
Indeed, if $n^*$ did not exist, then $S(t,S_0)$ would 
belong to the open set $Q\cap \textrm{int}(\R^n_{\geq 0})$ for all $t>0$. In that set, the function $V$ 
is increasing along solutions as we have remarked earlier, and thus $\lim_{t\rightarrow +\infty} V(S(t,S_0))> V(S_0)>0$, contradicting that $\liminf_{t\rightarrow +\infty} V(S(t,S_0))=0$ since 
$S(t_n,S_0)\rightarrow S_e$ and $V(S_e)=0$. A similar argument shows that 
$\tau_n\rightarrow +\infty$ as $n\rightarrow +\infty$.

For all $n>n^*$ and $t\in (\tau_n, t_n)$, we have that 
$S(t,S_0)\in Q\cap \textrm{int}(\R^n_{\geq 0})$ and hence $V$ is increasing along this part of the solution. In particular, for all $n>n^*$:
$$
0<V(S(\tau_n,S_0))<V(S(t_n,S_0)),
$$
where the quantity on the right tends to zero as $n\rightarrow +\infty$. Consequently, 
$$
V(S(\tau_n,S_0))\rightarrow 0\textrm{ as } n\rightarrow +\infty.
$$
By the definition of $V$, it follows that $V(S(\tau_n,S_0))=c p_n$, where $S(\tau_n,S_0)=[*,p_n]'$ for all 
$n>n^*$. Since $c \gg 0$, it follows that $p_n\rightarrow 0$ as $n\rightarrow +\infty$.

By continuity of solutions, the sequence $S(\tau_n,S_0)$ belongs to $Q^c$, 
the complement of $Q$, which is a 
closed set. Since the sequence $S(\tau_n,S_0)$ is bounded, we can 
pass to a subsequence if necessary, and  
assume that $S(\tau_n,S_0)\rightarrow p\in Q^c$ as $n\rightarrow +\infty$. By the previous paragraph, 
$p$ must belong to $L_{\Sigma}$, and $p$ clearly belongs to 
$\omega(S_0)$ as well. Finally, since $p\in Q^c$, there holds that $p\neq S_e$, which establishes the claim.

On the other hand, let 
$C:=  \omega(S_0)\cap \textrm{cl} (L_\Sigma )$. By definition $C$ is a compact set. 
We wish to show that $C$ is invariant. First, any solution starting in $C$ exists for all $t\in \mathbb{R}$ 
since it belongs to the compact invariant set $\omega(S_0)$. 
Moreover, by $(\ref{infinitesimal})$ and denoting by $\sigma(t)$ the $\sigma$-component of any such solution at time $t$, it holds $\sigma(t) = 0$ for all $t \in \mathbb{R}$, thus showing that 
such solutions remain in $\textrm{cl}(L_\Sigma)$ for all $t \in \mathbb{R}$. It follows 
that $C$ is indeed invariant. 
Hence, Corollary $\ref{GAS2}$, applied with
$C $ equal to the closure of $\omega(S_0)\cap L_\Sigma$, 
and $x_0 = S_e$, implies that $C = \{ S_e \}$, which contradicts the fact that
$p$ belongs $C$.
\epr

\section{Example: Apoptosis Regulation Pathway}
We consider the model proposed in \cite{eissing} of a network responsible for the regulation of Apoptosis (cell death).
This comprises the reactions listed below:
\be{apoptosis}
\begin{array}{rcccl}
C_8^{\star} + C_3 &\rightarrow & C_8^{\star} + C_3^{\star} \\
C_8 + C_3^{\star} &\rightarrow & C_8^{\star} + C_3^{\star} \\
C_3^{\star} + IAP & \leftrightarrow & C_3^{\star}-IAP &\rightarrow & \emptyset \\
C_3^{\star} + IAP & \rightarrow & C_3^{\star} &\rightarrow & \emptyset \\
C_8^{\star} + BAR & \leftrightarrow & C_8^{\star}-BAR & \rightarrow & \emptyset \\ 
 \emptyset & \rightarrow & IAP & \rightarrow & \emptyset \\
 \emptyset & \rightarrow & BAR & \rightarrow & \emptyset \\
\emptyset & \rightarrow & C_8 & \rightarrow & \emptyset \\
\emptyset & \rightarrow & C_3 & \rightarrow & \emptyset \\
& & C_8^{\star} & \rightarrow & \emptyset 
\end{array}
\ee
\begin{figure}
\centerline{
\includegraphics[width=7cm]{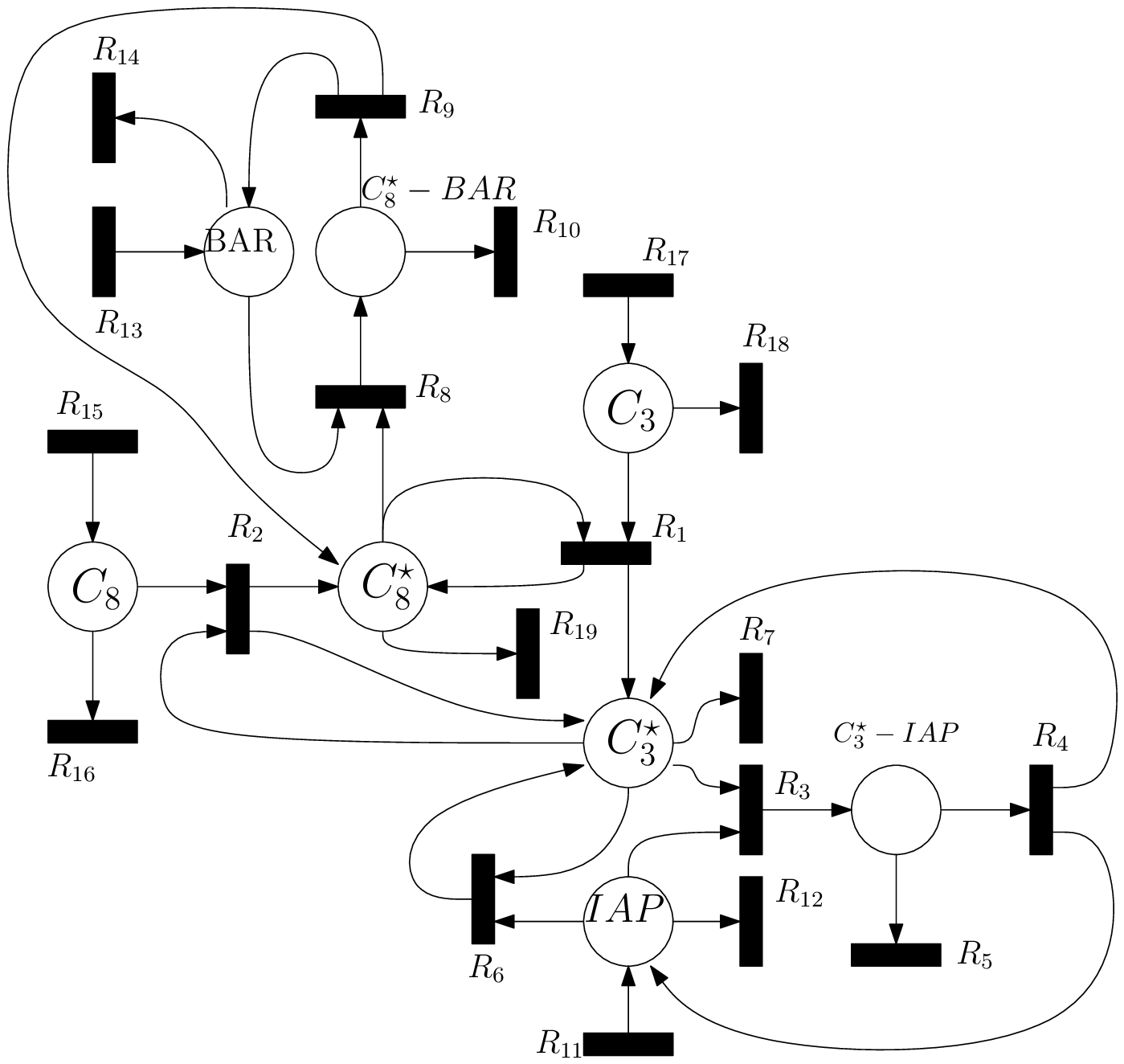}}
\caption{Apoptosis Network}
\label{apop}
\end{figure}
Due to the presence of many inflows as well as outflows (degradation of molecules), the network does not exhibit conserved moieties; in the language
of Petri Nets, there are no $P$-semiflows. 
Computation of $T$-semiflows, yields besides the obvious ones, $R_3+R_4$, $R_8+ R_9$, $R_{13} + R_{14}$, $R_{15}+ R_{16}$, $R_{11}+ R_{12}$, $R_{17}+R_{18}$, entailed by the presence of reversible reactions,  the following vectors:\\
\begin{center}
\begin{tabular}{c}
\hline
 $R_{6} + R_{11}$ \\ $R_{2} + R_{15} + R_{19}$ \\  $R_{1}+ R_{7} + R_{17}$ \\ $R_{1}+ R_{3} + R_{5} + R_{11} + R_{17}$ \\ 
 $R_2 + R_8 + R_{10} + R_{13} + R_{15}$  \\
 \hline
\end{tabular} 
\end{center}
Notice that every reaction belongs to at least one $T$-semiflow, hence the
network is consistent, and fulfills the necessary condition for persistence. 
Thus, it is interesting to look for sufficient conditions for persistence.
Analysis of Input-Output stability of the network is made possible by the
presence of outflows in every chemical species. 
Indeed, by letting
\be{Vlyap}
\begin{array}{rl}
V =&
[C_8]  + [C_8^{\star}]  + [C_3] + [C_3^{\star}]  + [IAP] \\  &+ [BAR]  + [C_8^{\star}-BAR]  + [C_3^{\star}-IAP] 
\end{array}
\ee
and taking its derivative with respect to time yields:
\be{vder}
\begin{array}{rl}
\dot {V}  & \leq  (k_{17} + k_{15} + k_{13} + k_{11} ) - k_{16} [C_8]  - k_{19} [C_8^{\star}] \\ & - k_{18} [C_3] - k_7 [C_3^{\star}]  
- k_{12} [IAP]   - k_{14} [BAR]  \\ & - k_8 [C_8^{\star}-BAR]  - k_5 [C_3^{\star}-IAP]  
\leq  K_{in} - K_{out} V
\end{array}
\ee
where the last inequality follows by letting:
\be{constants}
\begin{array}{rcl}
K_{in} &=& (k_{17} + k_{15} + k_{13} + k_{11} )  \\
K_{out} &=& \min \{k_{16}, k_{19}, k_{18}, k_7, k_{12},k_{14}, k_8, k_5 \}.
\end{array}
\ee
Hence, $\limsup_{t \rightarrow + \infty} V(t) \leq K_{in} / K_{out} < + \infty$ which proves input-output stability of the network
(this is actually true even for time-varying inflows, in which case the $\sup$ norm of $K_{in} (t)$ is needed in constructing the estimate).
To further analyze the network it is useful to investigate the presence of siphons. Indeed, there exists one minimal siphon:
$\Sigma = \{ C_3^{\star}, C_8^{\star}, C_3^{\star} - IAP, C_8^{\star}-BAR \}$. 
Due to the absence of $P$-semiflows this is indeed a critical siphon. 
Let $\sigma =[ C_3^{\star}, C_8^{\star}, C_3^{\star} - IAP, C_8^{\star}-BAR ]'$ be the concentration vector of species belonging to
$\Sigma $ and $\zeta = [ IAP, C_8, C_3, BAR ]'$.
Writing down explicit equations for $\dot {\sigma }$ yields:
\be{sigmaeq}
\dot {\sigma } =
\left [ \begin{array}{c}
- k_7 \sigma _1 - k_3 \sigma _1  \zeta _1 + k_1 \zeta _3  \sigma _2 + k_4 \sigma _3 \\
- k_{19} \sigma _2 - k_8 \sigma _2  \zeta _4 + k_9 \sigma _4 + k_2 \zeta _2  \sigma _1 \\
- (k_4+k5) \sigma _3 + k_3 \sigma _1  \zeta _1 \\
- (k_9 + k_{10} ) \sigma _4 + k_8 \sigma _2  \zeta _4
\end{array} \right ]
 \ee 
Taking the Jacobian of the vector-field for $\sigma =0$, yields:
\[
J=\left [ \begin{array}{cccc} -k_7 - k_3 \zeta _1 &  k_1 \zeta _3 & k_4 & 0 \\
      k_2  \zeta _2        & -k_{19} - k_8 \zeta _4 & 0 & k_9 \\
 k_3 \zeta _1 & 0 & -k_4 - k_5 & 0 \\
      0 & k_8 \zeta _4 & 0 & -k_9 - k_{10}
     \end{array}
     \right ] 
\]  
The species not included in $\Sigma $, on the other hand, evolve according to the following simple equations (once each variable in $\Sigma $ is set to
$0$):
\be{outofsigma}
\begin{array}{rcl}
\dot {\zeta }_1 &=& k_{11} - k_{12} \zeta _1 - k_3 \sigma _1 \zeta _1 - k_6 \sigma _1 \zeta _1 + k_4 \sigma _3\\
\dot {\zeta }_2 &=& k_{15}-k_{16} \zeta _2 - k_2 \sigma _2 \zeta _2\\
\dot {\zeta _3} &=& k_{17}-k_{18} \zeta _3 - k_1 \sigma _2 \zeta _3\\
\dot {\zeta _4} &=& k_{13} - k_{14} \zeta _4 - k_8 \sigma _2 \zeta _4 + k_9 \sigma _4
\end{array}
\ee 
which letting $\sigma =0$ yields the following simple globally asymptotically stable embedded system:
\be{outofsigma2}
\begin{array}{rcl}
\dot {\zeta }_1 &=& k_{11} - k_{12} \zeta _1 \\
\dot {\zeta }_2 &=& k_{15}-k_{16} \zeta _2 \\
\dot {\zeta _3} &=& k_{17}-k_{18} \zeta _3 \\
\dot {\zeta _4} &=& k_{13} - k_{14} \zeta _4 
\end{array}
\ee 
whose equilibrium is located at $[ k_{11}/k_{12}, k_{15}/k_{16}, k_{17}/k_{18}, k_{13}/k_{14} ]$. 
Hence, we have a boundary equilibrium at $[ k_{11}/k_{12}, k_{15}/k_{16}, k_{17}/k_{18}, k_{13}/k_{14} , 0,0,0,0]$.

Applying Lemma $\ref{hof}$ from the Appendix leads to the following.
\bl{hu}
Let $\zeta_i>0$ be arbitrary. Then $J$ is Hurwitz if and only if its determinant is positive.
\el
\bpr

Necessity is trivial. To prove sufficiency we calculate the 4 leading principal minors:
\begin{eqnarray*}
M_1&=&-k_7-k_3\zeta_1\\
M_2&=&(k_7+k_3\zeta_1)(k_{19}+k_8\zeta_4)-k_1k_2\zeta_2\zeta_3\\
M_3&=&-(k_4+k_5)\left(k_7(k_{19}+k_8\zeta_4)-k_1k_2\zeta_2\zeta_3 \right)\\
&&-k_3k_5\zeta_1 (k_{19}+k_8\zeta_4)\\
&=&-(k_4+k_5)M_2+k_4k_3\zeta_1(k_{19}+k_8\zeta_4)\\
M_4&=&-(k_9+k_{10})M_3-k_9\left(k_7(k_4+k_5)+k_3k_5\zeta_1 \right)k_8\zeta_4
\end{eqnarray*}
It follows that:
$$
\textrm{det} (J)=M_4>0\;\; \Rightarrow \;\; M_3<0\;\; \Rightarrow \;\; M_2>0,
$$
and $M_1<0$ is immediately clear. Thus, by Theorem $\ref{hof}$ in the Appendix, $J$ is Hurwitz whenever its determinant is positive.

\epr

Consequently, if the determinant of $J$ is positive, then the CRN is not persistent. On  the other hand, 
if the determinant of $J$ is negative, then it follows from Theorem \ref{conservativenetworkpersistence} 
that the CRN is persistent (notice that $J$ is irreducible).

\section{Conclusions}

New checkable criteria for persistence of chemical reaction networks have been
proposed, which apply even when kinetic rates are time-dependent.
These ``time-dependent'' rates may represent inflows and outflows, as well as
the effect of external inputs.
Finally, the case when critical siphons are present is also studied.
As an illustration, a hypoxia and an apoptosis network are analyzed.

\appendix
\section{Appendix}
In this Appendix we state and prove a few results used in the  proof of Theorem 
$\ref{conservativenetworkpersistence}$. 

\bl{aux1}
Assume that $g:\R^n \times \R^m\rightarrow \R^m$ is a $C^1$ vector function in some open set 
containing $\R^{n+m}_{\geq 0}$, and that it maps $(x,y)$ to $g(x,y)$.

Suppose that there is some $c \gg 0$ such that:
\begin{enumerate}
\item $g(x,0) = 0$  for all $x \gg 0$.
\item  $c \partial g/ \partial y({\bar x},0) \gg 0$  at some ${\bar x} \gg 0$.
\end{enumerate}
Then $c g(x,y) > 0$ for all $x$ in a neighborhood of ${\bar x}$ and all $y \gg 0$ 
in a neighborhood of $y=0$.
\el

\bpr
For each $(x,y)$, consider $m(t) = g(x,ty)$ as
a function of $t$ in $[0,1]$.  By the Fundamental Theorem of Calculus, and using $g(x,0)=0$:
$$
g(x,y) = g(x,0) + \int_0^1 (dm/dt)(t) dt =F(x,y) y,
$$
where $F(x,y) = \int_0^1 (\partial g/ \partial y) (x,ty) dt$.

Pick a convex neighborhood $Q$ of $({\bar x},0)$, where $c \partial g/ \partial y (x,y) \gg 0$ 
for all $(x,y)$ in $Q$.
By convexity, also $c \partial g/ \partial y(x,ty) \gg 0$ for all $t$ in $[0,1]$ and all $(x,y)$ in $Q$; 
therefore $c F(x,y) \gg 0$ for all $(x,y)$ in $Q$, 
and hence $c g(x,y) > 0$ if $(x,y)$ is in $Q$ and $y \gg 0$.
\epr

\bl{GAS}
Suppose that $C$ is a compact, invariant set of a 
dynamical system generated by a system ${\dot x}=F(x)$ defined on some set $X$  
in $\R^n$, and that $x_0$ is a point such that:
\begin{enumerate}
\item  For every compact, forward invariant subset $D$ of $C$, $x_0$  belongs to $D$.
\item For every $x$ in $C$ different from $x_0$, $x_0$ does not belong to $\alpha(x)$.
\end{enumerate}
Then, $C = \{x_0\}$.
\el
\bpr
Suppose that there is some $x$ in $C$ which is different from $x_0$.
Consider $D = \alpha(x)$.  
Since $C$ is compact and invariant, $D$ is nonempty, compact, and invariant.
Since $D$ is, in particular, forward invariant, it follows from 1 that
$x_0 \in D$.
A contradiction is obtained because 2 asserts that $x_0$ does not belong to $D$.
\epr

\bc{GAS2}
Suppose that $C$ is a compact, invariant set of a dynamical system 
generated by a system ${\dot x}=F(x)$ defined 
on some set $X$ in $\R^n$, and 
that $x_0$ is globally asymptotically stable. Then, $C = \{x_0\}$.
\ec
\bpr
Denote the flow on $C$ generated by the system by $\phi(t,x)$. 
We verify properties 1 and 2 from Lemma $\ref{GAS}$.
Property 1 is clear because of attractivity of $x_0$.
Property 2 follows from stability of $x_0$. Indeed, pick a neighborhood $U$
of $x_0$ which doesn't contain $x$, and a neighborhood $V\subset U$ of $x_0$
such that trajectories starting in $V$ cannot exit $U$ in positive time.
If $x_0$ belongs to $\alpha(x)$, then there is some $t>0$ such that $\phi(-t,x)\in V$. 
Then  $x = \phi(t,\phi(-t,x))\in U$, a contradiction.
\epr

In the example of the regulation pathway for apoptosis, the following basic result from 
\cite{hofbauer} is used.
\bl{hof}
A Metzler matrix is Hurwitz if and only if its leading principal minors alternate in sign, with the first 
one being negative.
\el

\end{document}